# ARQ Security in Wi-Fi and RFID Networks

Mohamed Elsabagh*, Yara Abdallah †, Moustafa Youssef‡, and Hesham El Gamal †
* Wireless Intelligent Networks Center (WINC)
Nile University, Cairo, Egypt
Email: mohamed.elsabagh@nileu.edu.eg
‡ Department of Computer Science and Engineering
Egypt-Japan University of Science and Technology, Alexandria, Egypt
Email: moustafa.youssef@ejust.edu.eg
† Department of Electrical and Computer Engineering
Ohio State University, Columbus, USA
Email: abdallah.22@osu.edu, helgamal@ece.osu.edu

*Abstract*—In this paper, we present two practical ARQ-Based security schemes for Wi-Fi and RFID networks. Our proposed schemes enhance the confidentiality and authenticity functions of these networks, respectively. Both schemes build on the same idea; by exploiting the statistical independence between the multipath fading experienced by the legitimate nodes and potential adversaries, secret keys are established and then are continuously updated. The continuous key update property of both schemes makes them capable of defending against all of the passive eavesdropping attacks and most of the currently-known active attacks against either Wi-Fi or RFID networks. However, each scheme is tailored to best suit the requirements of its respective paradigm. In Wi-Fi networks, we overlay, rather than completely replace, the current Wi-Fi security protocols. Thus, our Wi-Fi scheme can be readily implemented via only minor modifications over the IEEE 802.11 standards. On the other hand, the proposed RFID scheme introduces the first provably secure low cost RFID authentication protocol. The proposed schemes impose a throughput-security tradeoff that is shown, through our analytical and experimental results, to be practically acceptable.

## I. INTRODUCTION

Recent research on wireless information theoretic security has been largely motivated by the seminal works of Wyner on the wiretap channel [1], [2]. In those works, two users want to exchange secure messages over noisy channels, in the presence of an eavesdropper that is also impaired by a noisy channel. [1] shows that a non-zero secrecy capacity (the maximum achievable secure rate) is achievable, **without any assumptions on the computational power available to the eavesdropper or the presence of a private key that is shared between the legitimate users**, if the eavesdropper's channel is a degraded version of the legitimate one. This result was later extended to the non-degraded channel scenario in [3]. The effect of multipath fading on the secrecy capacity was studied in several recent works (e.g., [4], [5]). These works showed that the secrecy capacity could be enhanced if the secure message is judicially distributed across different channel fading realizations. Building on this principle, in our previous works [6], [7], we developed a framework for sharing keys over multipath fading channels using ARQ feedback. Under the assumption of a public and error-free ARQ feedback channel, the achievability of non-zero key rates, even when the eavesdropper is enjoying relatively better channel conditions (a higher average signal-to-noise ratio (SNR)) than the legitimate receiver, was established.

Inspired by the aforementioned results, we develop the first ARQ-based security protocols for two popular, and severely flawed, wireless communication paradigms: Wi-Fi and RFID. More specifically, the key idea enabling our design is the *opportunistic secrecy principle* which allows for exploiting the statistical independence between the multipath fading experienced by the legitimate nodes and adversaries to enhance the confidentiality and the authenticity functions of the underlying protocols.

For Wi-Fi networks, through a uniform treatment of the three existing Wi-Fi security protocols, we further extend our previous work [8] to a *security overlay* that provides information theoretic confidentiality guarantees to complement the security guarantees offered by the underlying protocols. By judiciously using the existing Automatic Repeat reQuest (ARQ) protocol in the IEEE 802.11 standard, our overlay relies on the opportunistic secrecy principle to establish a secret key that is shared only by the legitimate nodes; as described in the sequel. Remarkably, this goal is achieved through only minor modifications in the MAC layer. Our experimental results, obtained from our prototype implementation using the Madwifi driver, demonstrate the ability of our approach to defend against all of the passive Wi-Fi attacks and all of the currently known active attacks, at the expense of a minor loss in throughput and a small increase in link setup time, that are shown to be practically acceptable.

Moreover, we propose an ARQ-based provably secure RFID authentication protocol. Our novel scheme employs a low cost ARQ-based approach to securely share a secret key between the RFID reader and the tag and perform mutual authentication, in the presence of a passive eavesdropper. Our analytical results show the ability of our scheme to defend against passive attacks, demonstrating the attained tradeoff between secrecy and throughput. In addition, several key active attack techniques are shown to be inhibited.

The remainder of this paper is organized as follows. Section II gives an overview of the existing security protocols and their corresponding attacks in both Wi-Fi and RFID networks. In Sections III and IV, we provide the details of our proposed schemes, as well as a rigorous analysis of their security and performance in Wi-Fi and RFID networks, respectively. Finally, Section V offers some concluding remarks.

## II. BACKGROUND

### A. Wi-Fi Security

Since the emergence of the IEEE 802.11 standard, there have been numerous efforts to design provable and user friendly security protocols. These efforts resulted in three major security protocols, namely, the WEP (Wired Equivalent Privacy), WPA and WPA2 (Wi-Fi Protected Access) protocols. In general, the security functions of those protocols could be separated into three layers, namely, an authentication layer, an access control layer and a WLAN layer [9]. The processes involved with encrypting and decrypting frames are found in the WLAN layer solely (WEP, the Temporal Key Integrity Protocol (TKIP), and the Counter Mode with Cipher Block Chaining Message Authentication Code Protocol (CCMP)). All of those protocols attempt to use a **fresh key** for securing each transmitted frame using special security parameters. In the WEP protocol, a 24-bit value, called the Initialization Vector (IV), is generated and then combined with a secret root key. The result of this process is used for encrypting each frame. TKIP makes use of a similar 48-bit value, called TKIP Sequence Counter (TSC), while CCMP generates the Packet Number (PN), which is of length 48 bits as well. Since those parameters will be needed for decryption at the receiver, they are sent, **in-the-clear**, in the transmitted frame's header. For the purpose of this paper, we use the symbol $V$ to refer to WEP's IV, TKIP's TSC or CCMP's PN.

We now give a summary on the confidentiality-related attacks on the three Wi-Fi security protocols. Borisov, Goldberg, and Wagner first reported WEP design failures in [10]. Later, the first key recovery attack against WEP (the FMS attack) was presented by Fluhrer, Mantin and Shamir [11] using some weaknesses of the RC4 Key Scheduling Algorithm. The KoreK chopchop attack attempted at breaking WEP using the CRC32 checksum (the ICV test) [12]. KoreK also presented another large group of attacks [13]. A rather efficient iterative algorithm that recovers the WEP key was later proposed by Klein in [14]. On the other hand, the Bittau attack made use of the fragmentation support of IEEE 802.11 to break WEP [15]. Finally, Pyshkin, Tews, and Weinmann presented more enhancements to the Klein attack by using ranking techniques [16]. At the moment, this recent attack is considered to be the most powerful attack against WEP.

In 2004, weaknesses in the temporal key hash of TKIP were shown [17]. An attacker could use the knowledge of a few keystreams and TSCs to predict the Temporal Key and the MIC Key used in TKIP. Later in 2008, Tews and Beck [18] made the first practical attack against TKIP. In this attack, an attacker can recover the plaintext of a short packet and falsify it within about 12-15 minutes. Based on this attack, in 2009, a practical falsification attack against TKIP was proposed [19], which shortens the attack time to about one minute. Finally, CCMP arguably provides the most robust form of security nowadays. However, a weakness in the nonce construction mechanism in CCMP was recently discovered [20]. A predictable PN in CCMP was shown to decrease the effective encryption key length from 128 bits to 85 bits [20].

In summary, the previously mentioned statistical attacks rely on collecting a large number of ciphertext along with the corresponding security parameters (e.g. the IV in WEP) *which are sent in-the-clear*, whether through passive eavesdropping or innovative active techniques. The remaining attacks may not need a large number of eavesdropped frames. They, however, critically depend on the availability of the frames' security parameters as plaintext. As detailed in the sequel, our proposed overlay transforms those security parameters into a secret key that is only shared by the legitimate nodes, and thus, defends against those attacks.

### B. RFID Security

Typically, an RFID system consists of three main components: an interrogator (reader), a transponder (tag), and a back-end database (server)[1]. Low-cost RFID tags have limited computational capabilities, which makes them incapable of performing sophisticated cryptographic operations. Since tags respond to interrogation without alerting the user, users have become concerned with the security and privacy ramifications associated with RFID. This lead to the need to achieve mutual authentication between the reader and the tag, so that sensitive information are only communicated between authentic parties.

A variety of RFID schemes that achieve mutual authentication between readers and tags were proposed. The work of Peris-Lopez et al. (e.g., [21]) introduced multiple mutual authentication protocols that depend on using only bitwise operations. However, they were shown to be insecure against several kinds of passive attacks [22]. Later on, it was shown that secure mutual authentication cannot be achieved by using only bitwise operations [23]. To achieve enhanced security levels, schemes that depend on hash-locks were proposed. The basic idea behind these schemes is to achieve privacy preservation by hashing the $ID$ of the tag. Typically, a random nonce $R$ is generated, concatenated with the $ID$, and the hash of the overall string is sent to the reader. In all these schemes, the reader is required to perform an exhaustive search over the back-end database in order to find a matching tag $ID$. A wide set of symmetric key solutions were also proposed, in addition to other solutions that depend on NP-hard problems, such as the learning parity with noise (LPN) problem as in [24]. Although hash-lock schemes are designed to provide high security levels, an assumption is made here that the attacker (either active or passive) has no access to the tags $ID$ space, and thus will not be able to unlock the hash and uncover the $ID$ of the tag.

---

[1]Throughout the sequel, we make no distinction between the reader and the server.

To sum up, the main bottom layer of all these RFID protocols is the supposition that the employed cryptographic primitives are *computationally secure*. That is, it is computationally hard for a attacker to recover the secret information, without knowing the secret key. In fact, in schemes that employ hash-locks or privacy preservation, the reader behaves like a brute force attacker, with an additional advantage which is that the brute forcing space is reduced to the actual number of tags in the system. These schemes, however, do not prevent a passive eavesdropper from recovering secret information by, for example, statistical analysis or even brute forcing the secret key. That says, these schemes are not provably secure. Recently, a step towards information theoretic secrecy for RFID is made by Alomair et al. in [25], in which they transmit a random nonce from the reader to the tag using only modulo $p$ addition and multiplication operations, where $p$ is a large prime number. However, their scheme is based on the assumption that recovering the value of a nonce by eavesdropping enough protocol runs is impractical in RFID, which questions the actual resiliency of the scheme to passive key disclosure and statistical analysis attacks.

Based on information theoretic principles, we provide the first provably secure scheme for RFID. We show that through employing a simple ARQ-based approach, our proposed RFID scheme provides provable security without any limiting assumptions on the eavesdropper.

## III. ARQ Security for Wi-Fi Networks

### A. Protocol Description

Our proposed overlay is designed for Wi-Fi networks operating in infrastructure mode that may use any of the IEEE802.11 security protocols, i.e., WEP, TKIP or CCMP for encryption. The network is composed of one AP and $L$ clients, in the presence of one attacker. The AP and all clients follow the ARQ mechanism adopted in the IEEE 802.11 standard. In this paper, we assume disabled retransmissions[2]. In the proposed overlay, we transform the $V$ values of different frames into additional private keys that are shared among the legitimate nodes. The proposed protocol works in two phases. (a) The Initialization Phase: this is where the AP and each client agree on a secret key $V_0$ that is distributed across many ARQ epochs and then used to initialize the $V$ values for the encrypted data frames. This phase occurs before any data exchange is allowed. (b) The Data Exchange Phase: here the AP and the clients transmit or receive data frames (whether unicast or multicast) while their $V$ values are being continuously updated using our overlay.

For ease of presentation, we begin by using a simple three-node network model. In this network, Alice corresponds to one legitimate client, Bob corresponds to the AP and Eve is a malicious attacker. We later show how to extend our scheme to secure multicast flows. Each pairwise channel in this network is assumed to be a block erasure channel. At time slot $i$, we denote the the erasure probability of each channel by $\gamma_{XY}(i) \in [0, 1]$, where $X$ is the transmitter and $Y$ is the receiver for that channel. The erasure probabilities are assumed to be random variables which are correlated across users but independent and identically distributed over time.

[2]The analysis provided in this paper could be easily extended to the case of enabled retransmissions.

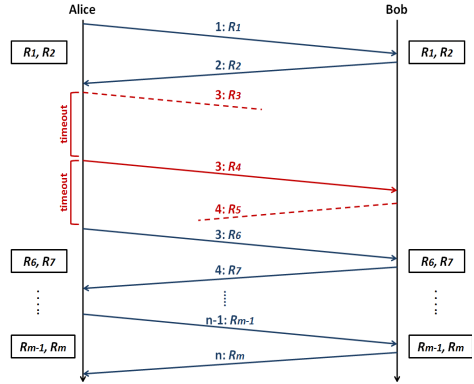

Fig. 1. The initialization phase.

*1) The Initialization Phase:* The initialization phase works as illustrated in Figure 1. First, Alice transmits an initialization frame, carrying a sequence number 1 and a uniformly chosen random number $R_1$, and starts a timer. Once Bob receives this frame, he replies with another initialization frame, carrying a sequence number 2, and another random number $R_2$. If Alice receives this frame before a timeout event occurs, she stores the pair $(R_1, R_2)$ for later use, and transmits another initialization frame with sequence number 3 and a new random number $R_3$. Otherwise (a timeout event occurs), Alice discards $R_1$, and transmits another initialization frame with sequence number 1 and a new random number $R_3$. On the other side, Bob keeps on responding to each initialization frame he gets with a sequence number incremented by one, and a newly generated random number. However, he stores only the last pair it has for any given sequence number. The process continues till Alice and Bob has stored $n$ initialization random values (exhausting $m$ trials, as shown in Figure 1). The length of each transmitted random number is 24 bits if WEP is used, or 48 bits otherwise. Finally, the secret key, $V_0$, is the modulo-2 sum of the random number pairs successfully received by **both** Alice and Bob.

*2) The Data Exchange Phase:* Right after initialization, our protocol works on updating the $V$ values, used to encapsulate each transmitted data frame. For unicast data flows, first consider the $i^{th}$ data frame to be securely transmitted, using any security protocol, from Alice to Bob. Alice starts by generating a random number (of length 24 if WEP is used, or 48 bits otherwise) referred to as the header-V, $V_h(i)$. The protocol must not use two consecutive equal header-V's. This property will be shown to be useful for defending against replay attacks. This value, $V_h(i)$, is put in the frame's security header, according to the specifications of the security protocol used. However, **unlike** the standards, the value used by our scheme in encapsulating the frame, denoted by $V_e(i)$, is the **modulo-2 sum** of the current header-V, $V_h(i)$, and all of the header-V's previously transmitted by Alice and successfully received by Bob. The update equation for $V_e$ is then

$$V_e(i) = \begin{cases} V_h(i) \bigoplus V_e(i-1), & \text{if } Q(i-1) = 1, \\ V_h(i) \bigoplus V_e(i-1) \bigoplus V_h(i-1), & \text{otherwise,} \end{cases} \quad (1)$$

where $Q(i) = 1$ if Alice received an ACK for the $i^{th}$ transmitted frame, $Q(i) = 0$ otherwise. This status is obtained through an ACK/Timeout detection module running at Alice. The initial value for this algorithm is set by the agreed-upon $V_0$ of the initialization phase, i.e., $V_e(0) = V_0$, while $V_h(0) = 0$. Similarly, when Bob receives the $i^{th}$ frame, he first extracts $V_h(i)$ from the security header, and then performs a check. If $V_h(i) = V_h(i-1)$, Bob discards the frame and treats it as a sign of a replay attack. If not, Bob attempts to decapsulate the frame with $V_d(i)$,

$$V_d(i) = V_h(i) \bigoplus V_d(i-1), \quad (2)$$

where $V_d(0) = V_0$. If decryption fails (an ICV failure occurs), this would be due to an erasure of the $(i-1)^{th}$ ACK. Bob then goes through another decryption attempt, after excluding $V_h(i-1)$ from the sum, i.e., with $V_d(i) = V_h(i) \bigoplus V_d(i-1) \bigoplus V_h(i-1)$. Another failure in decryption is treated as a sign of an attack and countermeasures could be invoked (the reason behind this will become clear in the security analysis to follow). Following this protocol, Alice and Bob perfectly agree on the $V$ values used for each frame. We avoid any mis-synchronization that could happen due to the loss of an ACK frame; without any additional feedback bits. The unicast flow from Bob to Alice could be secured in the same manner illustrated above.

Our scheme for multicast traffic goes as follows. Whenever a client subscribes to a multicast group, $g$, the AP sends a new random value, $V_g$, to every associated client that belongs to this group along with an ID for this $V_g$ value (the updates can be periodic or triggered based on group membership changes). Those values are transmitted to each client over its secure pairwise link with the AP, i.e., as *encrypted* frames. Once the AP makes sure that all clients in the group have received $V_g$, through individual ACKs, the AP uses this value to compute $V_{e_g}$, that will be used for encapsulating each upcoming multicast frame, within this group, i.e.,

$$V_{e_g}(i) = V_h(i) \bigoplus V_g, \quad (3)$$

where $V_h(i)$ is a random header-V as illustrated before. $V_h(i)$ and the ID of the used $V_g$ are sent in the header of the multicast frame. Similarly, for members of a particular multicast group $g$, a client uses the recovered information from the security header to compute $V_{d_g}(i)$ and decapsulate any multicast frame addressed to this group. Any failure in decryption (ICV test failure) is treated as a sign of attack. Finally, the AP should not use repeated $V_h$ values within the lifetime of a certain $V_g$. Similarly, whenever a client receives a multicast frame, it must check for this condition and treat repeated $V_h$'s as a sign of attack.

### B. Security Analysis

*1) Passive Eavesdropping:* The security of this protocol in the presence of a passive Eve directly builds on the results provided in [6], [8]. More specifically, as Eve becomes completely **blind** about $V_0$ if she misses **one** of the values constituting it, the probability that Eve correctly computes $V_0$ is

$$P_0 = \prod_{i \in \mathcal{A}} (1 - \gamma_{AE}(i)) \prod_{j \in \mathcal{B}} (1 - \gamma_{BE}(j)), \quad (4)$$

where $\mathcal{A}$ and $\mathcal{B}$ are the sets of time indices that correspond to the frames stored by Alice and Bob, respectively. $\gamma_{AE}(1), \ldots, \gamma_{AE}(n-1)$ denote the frame loss probabilities in the Alice-Eve channel whereas $\gamma_{BE}(2), \ldots, \gamma_{BE}(n)$ denote the frame loss probabilities in the Bob-Eve channel. All of those probabilities are random variables that are independently and identically distributed according to Eve's channels' distributions. Since the size of each of $\mathcal{A}$ and $\mathcal{B}$ is $n/2$. It is evident that, as $n$ increases, $P_0$ decreases and we achieve better security gains, at the expense of a larger delay in the initialization phase.

In our scheme, the collected traffic by a passive Eve becomes useful for any attack depending on Eve's ability to correctly compute $V_e$ for each captured frame. To achieve this, for unicast flows, Eve first has to correctly compute $V_0$, in the initialization phase between Alice and Bob. This happens with probability $P_0$ (as given in Eq. (4)). Afterwards, for each captured frame, Eve has to keep track of **all** the previously acknowledged data frames preceding that frame. Eve becomes, again, completely blind if she misses a **single acknowledged** frame. Based on this observation, we let $u$ denote the total number of data frames that Eve can correctly compute their $V_e$, i.e., the *useful* frames for Eve. If $\gamma_{AE} = \gamma_{BE} = \gamma_E$ for all time indices, the expected number of such frames is upper-bounded by[3]

$$\mathbb{E}[u] \leq \frac{\mathbb{E}[\gamma_E']^{n+1} - \mathbb{E}[\gamma_E']^{N+1}}{\mathbb{E}[\gamma_E]}, \quad (5)$$

where $\gamma_E' = 1 - \gamma_E$, $n$ is the total number of initialization frames constituting $V_0$ and $N$ is the unicast data session size. As shown in Eq. (5), a slight increase of the number of initialization frames results in a significant decrease in the number of useful frames for Eve in each session. Moreover, a passive Eve cannot make use of any of the multicast frames, as secure pairwise links are used to incorporate hidden and periodically-updated values into multicast $V_e$'s. This has a direct impact on the feasibility of many attacks, especially the statistical WEP attacks, e.g. [11], as those depend on collecting a large number of IVs ($V_e$'s in the ARQ overlay case) to run efficiently.

*2) Active Eavesdropping:* If Eve is active, she will be capable of injecting or replaying initialization or data frames. However, any replay or injection attempt would lead to a decryption failure at the legitimate recipients who will then treat this as a sign of an attack and countermeasures could follow. The most straightforward countermeasure is to change the keys of the whole network or of the attacked sessions. Still, for unicast flows, the history of the $V$ values built up thus far would still be used after invoking countermeasures.

---

[3]This bound is derived for the case in which Eve is perfectly capable of tracking the status of each transmitted data frame.

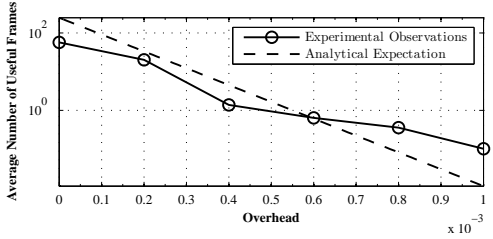

Fig. 2. The average number of useful frames at Eve. $\mathbb{E}[\gamma_{AB}] = 0.005, \mathbb{E}[\gamma_{BA}] = 0.009$ and $\mathbb{E}[\gamma_{AE}] = 0.004$.

### C. Experimental Results

Our experiments are conducted with a modified version of the Madwifi driver that has the capabilities of the proposed overlay. All of our testbed nodes are Dell Latitude D830 laptops that are equipped with Atheros-based D-Link DWL-G650 WLAN cards. All traffic is generated using Netperf [26].

*1) Security:* One-way traffic was generated between a client node (Alice) and the AP (Bob) in the presence of one eavesdropper (Eve). Eve's driver was equipped with the ARQ-based algorithms, i.e. Eve calculates $V_e$ for each frame based on the captured traffic. In our experiment, Eve had relatively better channel conditions, as compared to Bob[4]. We compared the $V_e$ values that Eve and Bob obtained for each frame, and calculated the number of useful frames for Eve (with different numbers of initialization frames). The results are reported in log scale in Figure 2. The data session size is taken to be 100000 frames. These results can be used to estimate the required time for Eve to capture a total of 1.5 million useful frames that is typically required to launch a combined form of the FMS and KoreK attacks ([27]). Under the original WEP operation, we assume that Eve needs 10 **minutes** to gather such traffic using passive eavesdropping only. Based on this estimate, using ARQ-WEP protocol extends the required average listening time for Eve to 1.24 **years**, using only an initialization overhead of 0.001. Finally, we note that under the ARQ overlay operation, Eve cannot use any active techniques to reduce the listening time. Consequently, any statistical WEP attack would becomes virtually impossible using our overlay. For TKIP and CCMP, the decreased number of useful frames at Eve hampers her ability to exploit the weaknesses that were previously discussed in Section II.

*2) Throughput:* Here we compare the performance of the proposed overlay with the baseline software implementations of WEP, TKIP, and CCMP in the Madwifi driver. To obtain a measure of performance if the proposed overlay was implemented in hardware, we also include the results of all

[4]We also report the results of a similar experiment in [8], where all nodes suffered from worse channel conditions.

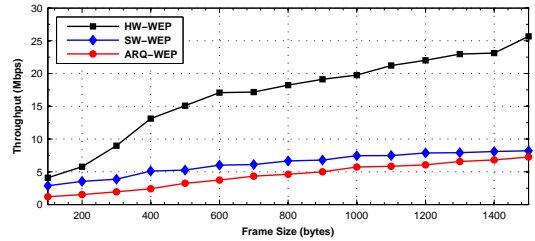
(a) WEP is used for encryption.

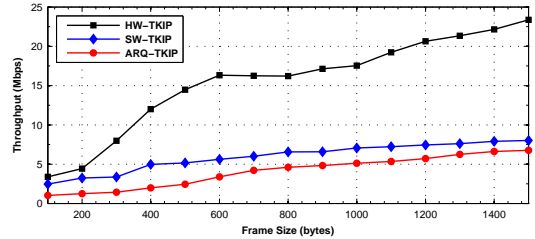
(b) TKIP is used for encryption.

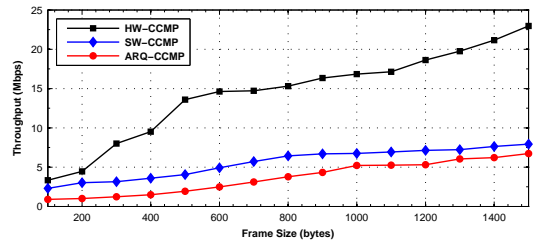
(c) CCMP is used for encryption.

Fig. 3. Network throughput for TCP flows with different security protocols.

hardware implementations. Figure 3 reports the aggregate network throughput for TCP flows, with different packet sizes, for WEP, TKIP, and CCMP. One can see that using the ARQ-based overlay on top of WEP (ARQ-WEP) results in a throughput degradation of 11.57% over the Madwifi software implementation of WEP (SW-WEP), for a packet size of 1500 bytes. The corresponding degradation for TKIP and CCMP is 15.61% and 15.26%, respectively. This quantifies the processing overhead of the additional ARQ-related operations. As the packet size increases, the overhead introduced by the ARQ-protocol decreases, as it is amortized over a larger packet size.

## IV. ARQ SECURITY FOR RFID NETWORKS

### A. Protocol Description

We follow the same three-node system model illustrated in Section III, where Alice corresponds to the reader, Bob to the tag, and Eve to the passive eavesdropper. Our scheme consists of three stages: 1) ARQ-based key sharing stage, 2) reader authentication stage, and 3) tag authentication stage. The three stages work together to defend against passive and active attacks on the system. Initially, the tag and the reader are assumed to share a pseudo-identity $IDS$, a secret key ($k \equiv k_1 || k_2 || k_3 || k_4 || k_5$), where $||$ denotes the string concatenation operation, along with the unique tag identifier $ID$. All identifiers and key components are assumed to be of length $\ell$ bits. We also assume that the reader initiates the protocol, and that both the tag and the reader are capable of computing a pseudo-random function (PRF) $f_s : \{0,1\}^{2\ell} \to \{0,1\}^{2\ell}$, where $s$ is a seed. The details of each stage are given below.

*1) ARQ Stage:* In this stage, the reader sends to the tag $m$ random numbers (drawn uniformly over the binary alphabet $\{0,1\}$). The tag responds in turn with a single ACK bit, or does not reply for a NACK. Similar to Section III-A1, the frames from the reader to the tag are sequenced in a manner that overcomes the loss of ACKs. The agreed upon key ($k' \equiv k'_1 || k'_2 || k'_3 || k'_4 || k'_5$), is distilled as the **modulo-2 sum** of the successfully acknowledged frames. The ARQ stage acts as a shield to the remaining two stages as shown next.

*2) Reader Authentication Stage:* This stage proceeds as follows.

a) The tag chooses a random nonce $N_T$ and transmits $[IDS \oplus k'_1, N_T \oplus k'_2]$ to the reader.
b) The reader chooses a random nonce $N_R$, computes $m_1 \equiv f_{k_1 \oplus k'_3}(N_T, N_R)$, and transmits $[N_R \oplus k_2 \oplus k'_4, m_1]$ to the tag. The reader then updates the stored pseudo-identity and the secret key to:
$IDS^{new} = IDS \oplus k_3 \oplus N_R$, and $k^{new} = k \oplus k'$.
c) The tag recomputes and checks $m_1$, and authenticates the reader or aborts.

In the next stage, the reader authenticates the tag in a similar manner.

*3) Tag Authentication Stage:*

a) The tag computes $m_2 \equiv f_{k_4 \oplus k'_3}(ID, N_R)$ and transmits $[ID \oplus k_5 \oplus k'_5, m_2]$ to the reader. The tag then updates the stored pseudo-identity and the secret key in the same way as in stage 2.
b) The reader recomputes and checks $m_2$, and authenticates the tag or aborts.

The security of our proposed scheme is analyzed in the following section.

### B. Security Analysis

*1) Passive Attacks:* Using our scheme, and similar to the results in Section III, Eve becomes completely **blind** about the distilled key if she misses a **single acknowledged** frame. In addition, she also blinds if she misses a **single ACK** bit, as she would not be capable of performing a brute force attack on the status of each frame. In fact, in typical RFID scenarios, the tag (Bob) will be in close proximity to the reader (Alice) than it is to Eve and, since the radiated signal power from the tag is, by orders of magnitude, smaller than the signal of the reader [28] which results in a higher probability for Eve to loose the tag responses rather than to loose the random frames sent by the reader. Hence, the secrecy outage probability in our case is given by

$$\Pr_{out1} = \prod_{i \in \mathcal{A}} (1 - \gamma_{AE}(i))(1 - \gamma_{BE}(i)). \quad (6)$$

where $\mathcal{A}$ is the set, with cardinality $n$, of the time indices corresponding to the frames successfully acknowledged by Bob. We here assume that the ACK responses, being of short length, occur within the same fading realization of their corresponding random number frames.

We investigate another interesting case in which Eve knows, a priori, the current $IDS$ of the tag (e.g., by interrogating the tag in a preceding protocol run). This would enable her to extract $k'_1$ (which is the prefix of the ARQ key) and to then exhaustively XOR in and out the sniffed random frames sent by the reader, until a matching prefix is found and thus, Eve becomes capable of overcoming the loss of ACK bits. With this added brute force capability at Eve, the probability of secrecy outage increases to

$$\Pr_{out2} = \prod_{i \in \mathcal{A}} (1 - \gamma_{AE}(i)), \quad (7)$$

In this case, if Eve looses $\ell$ ACKs, the complexity of her exhaustive search will blow up to $O(2^\ell)$. Hence, loosing a **single acknowledged frame, or $\ell$ or more ACKs** beats Eve back to brute force. On this account, the expected value of the probability of secrecy outage when Eve is computationally limited, such that she is capable of completing, in polynomial time, a brute force search of length less than $\ell$, is given by

$$\mathbb{E}\left[\Pr_{out3}\right] = \mathbb{E}\left[1 - \gamma_{AE}\right]^n \mathbb{E}\left[1 - \gamma_{BE}\right]^{n-\ell+1} \cdot \left[\sum_{i=0}^{\ell-1} \binom{n-\ell+i}{i} \mathbb{E}\left[\gamma_{BE}\right]^i\right]. \quad (8)$$

Figure 4 illustrates these bounds for $m = 30$, $\ell = 10$, a fixed $\mathbb{E}[\gamma_{AE}]$ of 0.02, and different values of $\mathbb{E}[\gamma_{BE}]$.

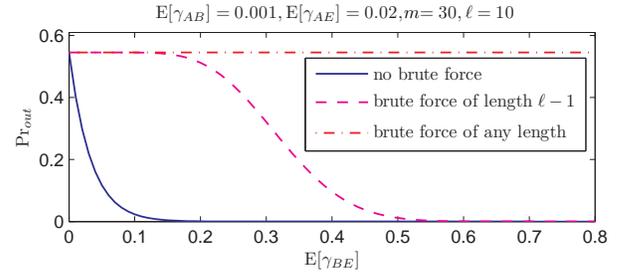

Fig. 4. Probability of secrecy outage for $m = 30$, $\ell = 10$, $\mathbb{E}[\gamma_{AB}] = 0.01$, $\mathbb{E}[\gamma_{AE}(i)] = 0.02$, and for different values of $\mathbb{E}[\gamma_{BE}]$.

Figure 5 captures the secrecy outage probability for various combinations of $\mathbb{E}[\gamma_{AE}]$ and $\mathbb{E}[\gamma_{BE}]$, for the two cases when Eve knows the $IDS$ of the tag and is computationally limited as in Eq. (8), and when she does not a priori know the $IDS$ of the tag as in Eq. (6), respectively. As shown, significant secrecy gains can be achieved even when the erasure probabilities for Eve's channels are as low as 0.01.

The choice of $m$ is essential for obtaining good security bounds, as well as reducing the communication overhead and the delay induced by the ARQ phase. Figure 6(a) shows the probability of secrecy outage for different values of $m$, a fixed $\ell = 10$, and a fixed erasure probability of 0.05 for Eve's channels. From the figure, it is clear that the secrecy outage probability **vanishes** as $m$ increases. Interestingly, our scheme grants a vanishing secrecy outage probability, even if Eve is experiencing a **better SNR** value than the legitimate parties. Figure 6(b) illustrates this for $\mathbb{E}[\gamma_{AB}] = 0.05$, while $\mathbb{E}[\gamma_{AE}] = \mathbb{E}[\gamma_{BE}] = 0.02$.

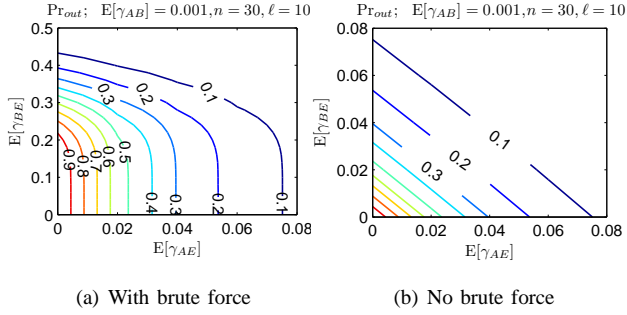

(a) With brute force

(b) No brute force

Fig. 5. Probability of secrecy outage for different combinations of $\mathbb{E}[\gamma_{AE}]$ and $\mathbb{E}[\gamma_{BE}]$. $\mathbb{E}[\gamma_{AB}] = 0.001$, $n = 30$, and $\ell = 10$.

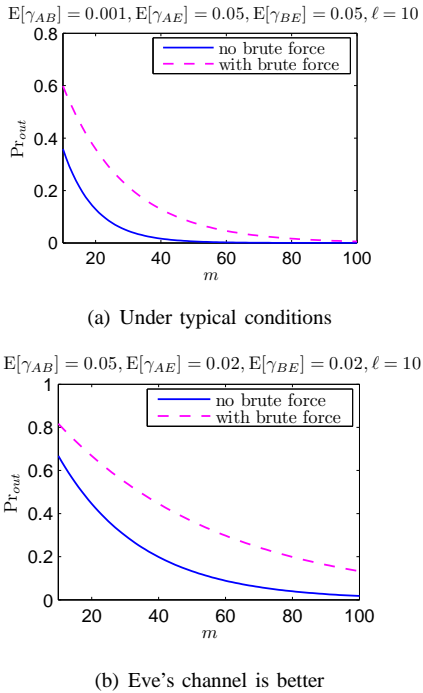

(a) Under typical conditions

(b) Eve's channel is better

Fig. 6. Probability of secrecy outage for different values of $m$.

From the aforementioned discussion, it becomes clear that our scheme is provably secure against passive attacks. The probability of secrecy outage, incorporated by our scheme, reaches its minimum when the eavesdropper cannot link a previously obtained $IDS$ to the victim tag. Additionally, updating the keys and the pseudo-identity of tags insures their **freshness**, which prevents tracking attacks as well.

*2) Key Disclosure Attacks:* In [29], a full disclosure attack against the protocol of Kim et al. [30] is illustrated. The attack depends on capturing multiple sessions with the same session nonce. The attack reveals the permanent secret key using a number of sessions that is extremely less than that needed for a brute force search. Since our scheme is secure against passive attacks, key disclosure by passive eavesdropping is thus defeated. In addition, an active adversary that is pretending to be a legitimate reader cannot initiate key disclosure attacks by interrogating the tag, since the tag transmits no sensitive information until *after* the reader authentication step. On the other hand, an active adversary may pretend to be a legitimate tag and may collect frames from the reader for a specific $IDS$. However, this can be easily detected by the reader as it will recognize so many failed or incomplete authentication attempts that are associated with the victim $IDS$, and therefore countermeasures can be employed.

*3) Synchronization Loss:* It is typical that the tag and reader update their keys only when the mutual authentication phase is completed. As this phase can be interrupted, either accidentally or intentionally, the tag and the reader may be desynchronized. To resolve that, we follow the same approach used in [25]: the reader stores the old keys and always updates its keys at step (2). When the tag returns, the reader will first try to authenticate it using the new keys. If this fails, the reader switches to the old keys and authenticates the tag, without giving it any further privileges, and then starts another protocol round with the tag using the new keys,. The tag is completely authenticated only if that second round succeeded. This way, the reader ensures that the tag is legitimate.

*4) Forward Secrecy:* Assuming that, at a time $t$, the adversary is given all the communication information between tags and readers along with all the information stored in a compromised tag. Forward secrecy requires the adversary not to be able to trace any past communication between the compromised tag and readers at any time $t' < t$ [31]. In our scheme, as tag's pseudo-identity and keys are refreshed; compromising a tag will not reveal any information about previously transmitted data. Thus, forward secrecy is achieved.

### C. Performance Analysis

We investigate the throughput-secrecy tradeoff governed by the number of achievable tag reads per second. A typical data rate of an HF reader of 106 kbps, and a tag read rate of 50 times per second [32] are assumed. We assume a key component length ($\ell$) of 10 bits, and that $\mathbb{E}[\gamma_{AE}] = \mathbb{E}[\gamma_{BE}] = 0.05$. Figure 7 illustrates the secrecy-throughput tradeoff associated with our scheme for different values of $m$. We ignore the delay caused by the three mutual authentication steps of the scheme. We are also assuming that the processing time required to XOR the random ARQ frames is negligible. The figure shows that our scheme is capable of achieving significant security gains, without great loss in throughput. For example, for $m = 100$, the scheme can achieve as low as $3 \times 10^{-5}$ secrecy outage probability, and approximately 20 tag reads per second.

Finally, table I summarizes the computational, communication, and storage overhead associated with our scheme.

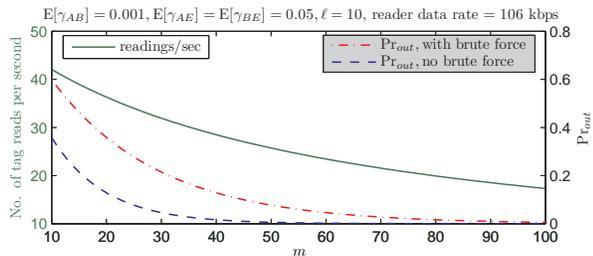

Fig. 7. Throughput-secrecy tradeoff for different values of $m$. We assume a typical HF reader data rate of 106 kbps, a key component length ($\ell$) of 10 bits, and that $\mathbb{E}[\gamma_{AE}] = \mathbb{E}[\gamma_{BE}] = 0.05$.

| | |
|---|---|
| No. of RNG operations | Reader : $m+1$ |
| | Tag : 1 |
| No. of PRF operations | Reader : 2 |
| | Tag : 2 |
| Protocol steps | $m+3$ |
| Memory size | $14\ell$ |

TABLE I
COMPUTATION, COMMUNICATION, AND MEMORY REQUIREMENTS.

## V. CONCLUSION

Inspired by information theoretic results, this paper developed two ARQ-based security schemes for Wi-Fi and RFID networks. The basic idea behind both schemes is to exploit the multipath fading characteristics of the wireless channel. The scheme developed for Wi-Fi protocols essentially enhances the confidentiality guarantees of Wi-Fi networks. When key management is a burden, which is the case in open access networks, our scheme opens a path for keyless confidentiality in these networks. Moreover, the principles introduced in this paper could be further extended to secure the currently deployed authentication protocols in Wi-Fi networks. Our second scheme, to the best of our knowledge, is the first low-cost RFID authentication scheme that provides provable security against passive attacks. Consequently, the currently known key attacks against RFID, such as key disclosure and tracking attacks, are totally inhibited. Finally, through our analytical and experimental results, we demonstrated the governing tradeoff between our secrecy gains and the network throughput and its practical appeal.